\documentclass[acmsmall,screen]{acmart}
\usepackage[normalem]{ulem}
\usepackage{microtype}

\usepackage{tikz}
\usetikzlibrary{positioning, arrows.meta}
\usepackage{amsmath}
\usepackage{amsthm}
\newtheorem{definition}{Definition}

\usepackage{algorithm}
\usepackage{algpseudocode}
\usepackage{booktabs}
\usepackage{xcolor}
\usepackage{listings}
\usepackage{pifont}
\newcommand{\DeepDiff}{\textsc{DeepDiff}}%
\usepackage{fancyvrb}
\usepackage{framed}
\usepackage{xcolor}
\usepackage{listings}
\usepackage{longtable}

\usepackage{subcaption}

\definecolor{codegray}{RGB}{245,245,245}
\definecolor{codegreen}{RGB}{0,110,0}
\definecolor{codeblue}{RGB}{0,0,150}

\lstdefinestyle{deepdiffcode}{
    basicstyle=\ttfamily\footnotesize,
    backgroundcolor=\color{codegray},
    keywordstyle=\color{codeblue}\bfseries,
    commentstyle=\color{codegreen},
    stringstyle=\color{red!60!black},
    numbers=left,
    numberstyle=\scriptsize\color{gray},
    numbersep=6pt,
    frame=single,
    rulecolor=\color{black!25},
    framesep=4pt,
    breaklines=true,
    breakatwhitespace=false,
    breakindent=0pt,
    postbreak=\mbox{\textcolor{gray}{$\hookrightarrow$}\space},
    columns=fullflexible,
    keepspaces=true,
    showstringspaces=false,
    tabsize=4,
    captionpos=b,
    aboveskip=0.7em,
    belowskip=0.7em,
    xleftmargin=1.5em,
    framexleftmargin=1.2em
}

\lstdefinestyle{deepdifftrace}{
    style=deepdiffcode,
    language={},
    numbers=none,
    basicstyle=\ttfamily\scriptsize
}

\begin{document}
\date{}
\title{\DeepDiff{}: Neuro-Dynamic Regression-Freedom Checking for Security Patch Validation}

\author{Laboni Sarker}
\authornote{Work was accomplished while the author was interning at Microsoft Research.}
\affiliation{%
  \institution{University of California, Santa Barbara}
  \city{Santa Barbara}
  \state{California}
  \country{USA}
}
\email{labonisarker@ucsb.edu}

\author{Shuvendu K Lahiri}
\affiliation{%
  \institution{Microsoft Research}
  \city{Redmond}
  \state{Washington}
  \country{USA}
}
\email{Shuvendu.Lahiri@microsoft.com}

\begin{abstract}
 Automatically validating AI-generated security patches remains a significant challenge. Eliminating a reported crash and passing existing test suites do not provide sufficient confidence that a patch correctly addresses the underlying vulnerability without introducing regressions.

 We present \DeepDiff{}, a technique for regression-freedom checking that differentially evaluates candidate patches against the original vulnerable program using non-crashing inputs that exercise the affected code paths. The expectation of regression-freedom is that a correct patch should eliminate the vulnerability triggering crash while preserving the intended behavior on non-crashing inputs. To support this evaluation, we introduce a dataset of correct and incorrect patch candidates for open-source memory-safety vulnerabilities from CyberGym. We explore different approaches for non-triggering test input generation to reach the patched code, a purely neural regression-freedom checker, and  a hybrid approach that combines dynamic analysis with neural techniques. Our experimental results demonstrate the effectiveness of regression-freedom checking in validating security patches while outlining the challenges that remain.

\end{abstract}
\maketitle

\section{Introduction}
\label{sec:intro}
Recent advances in large language models (LLMs) have enabled new approaches for discovering security vulnerabilities~\cite{sheng2025llms,li2025vulsolver,mythos_news,pesoli2026demystifying}, while simultaneously lowering barriers to vulnerability exploitation~\cite{wang2026exploitgym}. As the volume of discovered vulnerabilities continues to grow, manually developing patches for each vulnerability becomes increasingly challenging. This has motivated automated and AI-assisted approaches for security patch generation~\cite{han2025dependable,patUntrack,alrashedy2025leveraging,yang2026patch}.     However, the growing reliance on automatically generated patches necessitates effective techniques for rigorously evaluating their correctness~\cite{DBLP:journals/corr/abs-2606-04460,10659742,sajadi2025safe}.

Existing approaches for evaluating automatically generated security patches primarily focus on ensuring that the vulnerability is no longer triggerable in the patched program~\cite{wang2026cybergym,DBLP:journals/corr/abs-2606-04460,wei2025patcheval,lee2026sec}. Other approaches differentially compare generated patches against a golden patch to assess whether they are semantically equivalent~\cite{autoPatchBench,chen2025secureagentbench}. Although comparison with a golden patch provides a strong basis for establishing patch correctness, a golden patch is never available in real-world software patching scenarios. In the absence of a golden patch, it is therefore essential to evaluate whether the generated patch preserves the intended functionality of the program.

Existing approaches typically rely on developer-written tests or the project’s existing test suite for this purpose. However, these tests may not exercise the code affected by the patch and therefore may fail to expose functional divergence introduced by an incorrect patch. Consequently, a patch that builds successfully, passes the existing test suite, and prevents the vulnerability from being triggered may still be functionally incorrect. For example, a naive patch could eliminate the vulnerability by rejecting all inputs or by transforming inputs into a fixed non-vulnerable value, while also eliminating or altering the intended functionality~\cite{sarker2026quantitative,spider,black2018juliet,boland2012juliet}.

We address this gap with \emph{regression-freedom checking}, a validation criterion that uses the original vulnerable program as a behavioral reference, rather than relying solely on the existing test suite. A candidate patch (Definition~\ref{def:patch}) is considered plausible (Definition~\ref{def:valid}) if it eliminates the vulnerability-triggering crash while preserving the behavior of the original program on non-triggering inputs that the vulnerable program handles without crashing (Definition~\ref{def:benign}). A behavioral change on such an input indicates functional divergence, regardless of whether the patch also eliminates the original vulnerability. Moreover, the input that exposes this divergence serves as a concrete witness that can be inspected by developers to understand the behavior introduced by the patch.

We present \DeepDiff{}, a framework for regression-freedom checking that incorporates dynamic analysis, neural analysis, and a hybrid combination of the two (Section~\ref{sec:patchrx}). Given the original vulnerable program, its triggering input, and a candidate patched program, these techniques analyze both programs to identify functional divergences in the patched program relative to the non-triggering behavior of the original vulnerable program. The dynamic analyzer executes both programs on non-triggering inputs and compares their resulting observable behaviors to identify concrete witnesses of behavioral divergence. This approach provides executable evidence of divergence but requires building and maintaining an execution harness. In contrast, a neural analyzer reasons about the patch and its surrounding code context to identify witnesses without executing the programs, or to determine whether an observed divergence represents an unintended regression or an intentional behavior change. Moreover, a neural analyzer can generate non-triggering inputs to reach the patched code. Thus, neural analysis can support regression-freedom checking by generating inputs while reducing the need for an execution harness, but it may also produce spurious evidence. \DeepDiff{} combines the dynamic and neural components to leverage their complementary strengths.

We evaluate \DeepDiff{} on a dataset of correct and incorrect patches for open-source vulnerabilities drawn from CyberGym~\cite{DBLP:journals/corr/abs-2606-04460,cybergym_e2e_github}, comprising 630 candidate patches across 118 vulnerabilities from 40 projects. Of these, 385 candidates, including 134 developer-generated patches and 251 functionally divergent patches, form our regression-freedom evaluation corpus. Our results demonstrate that regression-freedom checking can identify functionally divergent patches that pass existing validation, while also highlighting an important remaining challenge: distinguishing unintended regressions from intentional behavior changes introduced by developers. Our results show that regression-freedom checking substantially improves detection of functionally divergent patches, increasing recall from 6\% to 64\%. Neuro-DeepDiff (Section~\ref{subsubsec:neural_deepdiff}) further achieves 79\% recall by generating candidate witnesses of behavioral divergence.

\textbf{Contributions:} Our contributions in this paper are as follows:
\begin{enumerate}
\item We introduce regression-freedom checking as a validation criterion for evaluating the functional correctness of security patches, using the original vulnerable program as the behavioral reference.

\item We develop and investigate dynamic, neural, and hybrid approaches for regression-freedom checking.

\item We construct a dataset of security patches, including developer-generated and functionally divergent patches, to support systematic evaluation of automated patch validation techniques.

\item We empirically evaluate the proposed regression-freedom checking approaches on security patch validation tasks.

\item We present case studies of developer-intended behavior changes on non-crashing inputs, highlighting the challenge of distinguishing intentional behavior changes from unintended functional regressions.
\end{enumerate}

\textbf{Organization.} The remainder of this paper is organized as follows. Section~\ref{sec:overview} provides an overview of regression-freedom checking through a motivating example. Section~\ref{sec:diffOracle} formalizes the concepts underlying regression-freedom checking. Section~\ref{sec:patchrx} presents the dynamic, neural, and hybrid instantiations of \DeepDiff{} for regression-freedom checking. Section~\ref{sec:implementation} describes our implementation. Section~\ref{sec:evaluation} presents our experimental evaluation and results. Section~\ref{sec:relwork} discusses related work, and Section~\ref{sec:conclusion} concludes the paper with possible future direction.

\section{Overview}
\label{sec:overview}
A security patch modifies an existing codebase to fix a vulnerability, and it should prevent the triggering input (proof of concept, PoC) from triggering that vulnerability. Eliminating the vulnerability, however, does not establish that the patch preserves the intended functionality of the original program: existing approaches~\cite{DBLP:journals/corr/abs-2606-04460} assess functional correctness with developer-written tests, which may not reach the patched code or assert the behavior it affects. We illustrate this gap with an example from the CyberGym benchmark.

The example is \texttt{opensc/arvo\_45552}, an out-of-bounds write in the \texttt{new\_file} function (Listing~\ref{lst:opensc-vulnerable}). In the vulnerable implementation, \texttt{map\_str2int} converts a name to a \texttt{df\_type} value and returns zero on success; the value is subsequently used as an index into \texttt{profile->df}.

\begin{lstlisting}[
    style=deepdiffcode,
    language=C,
    caption={Relevant vulnerable code in \texttt{new\_file}.},
    label={lst:opensc-vulnerable}
]
if (map_str2int(cur, name + 7, &df_type, pkcs15DfNames))
    return NULL;
file = init_file(SC_FILE_TYPE_WORKING_EF);
profile->df[df_type] = file;
\end{lstlisting}
Although \texttt{map\_str2int} may successfully parse a numeric value, the
vulnerable code does not verify that \texttt{df\_type} lies within the bounds
of the \texttt{profile->df} array. A triggering input is therefore a name
that parses successfully to a value outside the bounds of this array.

\paragraph{Developer Patch.} The developer patch in Listing~\ref{lst:opensc-correct} adds the required upper-bound check:
\begin{lstlisting}[
    style=deepdiffcode,
    language=C,
    caption={Developer patch.},
    label={lst:opensc-correct}
]
if (map_str2int(cur, name + 7, &df_type, pkcs15DfNames)
        || df_type >= SC_PKCS15_DF_TYPE_COUNT)
    return NULL;
\end{lstlisting}

\paragraph{Incorrect candidate.} An incorrect candidate patch in Listing~\ref{lst:opensc-incorrect} also prevents the crash on the
triggering input, but introduces two defects:
\begin{lstlisting}[
    style=deepdiffcode,
    language=C,
    caption={Incorrect candidate patch.},
    label={lst:opensc-incorrect}
]
if (!(map_str2int(cur, name + 7, &df_type, pkcs15DfNames))
        || df_type >= SC_PKCS15_DF_TYPE_COUNT - 1)
    return NULL;
\end{lstlisting}
First, \texttt{map\_str2int} returns zero when parsing succeeds, so negating its
result makes the candidate return \texttt{NULL} for every successfully parsed
name. This is also why the triggering input no longer crashes: it parses
successfully and is rejected by the negated condition, before the bound on
\texttt{df\_type} is evaluated. Second, the comparison against
\texttt{SC\_PKCS15\_DF\_TYPE\_COUNT - 1} is off by one and excludes the final
valid enumeration value.

Despite these defects, the incorrect candidate passes existing validation: it builds, passes the project's developer-written tests, and no longer crashes on the triggering input. Because every successfully parsed name is rejected, the out-of-bounds index is never reached, so the vulnerability is no longer triggerable, and no existing test asserts the behavior that the candidate changes. Existing validation therefore cannot distinguish the incorrect candidate from the developer patch.

The original vulnerable program, however, already distinguishes them. Consider a non-triggering input whose name parses successfully to an in-bounds \texttt{df\_type}. The vulnerable program handles this input without crashing: it creates the file and stores it in \texttt{profile->df[df\_type]}. The developer patch preserves this behavior, whereas the incorrect candidate returns \texttt{NULL} and the file is never created. The same holds for every valid name: the candidate removes the functionality instead of repairing it (Table~\ref{tab:opensc-behavior}). Each such input is a concrete witness of functional divergence, obtained without a golden reference patch and without any additional developer-written assertion. This is the observation behind regression-freedom checking that a patch should change the behavior of the vulnerable program on the triggering input, and preserve it on the non-triggering inputs that the vulnerable program already handles correctly.

\begin{table}[t]
    \footnotesize
    \centering
    \caption{Behavior of the vulnerable program, the developer patch, and the incorrect candidate for \texttt{opensc/arvo\_45552}.}
    \label{tab:opensc-behavior}
    \begin{tabular}{lccc}
        \toprule
        \textbf{Input} & \textbf{Vulnerable} & \textbf{Developer patch} & \textbf{Incorrect candidate} \\
        \midrule
        Triggering input      & Crash        & Rejected     & Rejected \\
        Valid in-bounds name  & File created & File created & Rejected \\
        \midrule
        Existing validation   & --           & Pass         & Pass \\
        \bottomrule
    \end{tabular}
\end{table}

\paragraph{Regression-freedom Checking.} Regression-freedom checking turns this observation into a validation criterion by treating the original vulnerable program as the specification of functional behavior on every input it handles without crashing. A candidate patch must change the behavior of the vulnerable program on the triggering input, but preserve it on non-triggering inputs. Unlike differential analysis against a ground-truth golden patch~\cite{autoPatchBench,chen2025secureagentbench}, the reference here is the vulnerable program itself, which is always available when a patch is being validated. Unlike the existing test suite, it requires no developer-written assertions or test oracles: any non-triggering input becomes a test whose expected behavior is given by the vulnerable program. Under this criterion, the developer patch in Listing~\ref{lst:opensc-correct} is plausible, while the candidate in Listing~\ref{lst:opensc-incorrect} is functionally divergent, even though existing validation accepts both. Thus, a divergence indicates that the patch requires further attention from the software developers as the patch is functionally divergent from the original behavior.

This example shows that comparing a candidate patch against the original vulnerable program on non-triggering inputs can distinguish a behavior-preserving repair from a functionally divergent patch that passes existing validation.

\section{Regression-Freedom Checking for Patch Validation}
\label{sec:diffOracle}
In this section, we formalize regression-freedom checking. We first define a vulnerable program, its triggering and benign (non-triggering) inputs, and a candidate patch. We then define what it means for a candidate patch to be functionally divergent or plausible, based on whether it preserves the behavior of the vulnerable program on benign inputs.

\begin{definition}
A \textbf{vulnerable program} $V$ is a program with a reported vulnerability. For a program $P$ and an input $x$, $\mathsf{Vuln}(P,x)$ denotes that executing $P$ on $x$ produces an observable crash or sanitizer violation.
\label{def:vulnprogram}
\end{definition}

\begin{definition}
An input $x_{\mathrm{PoC}}$ is a \textbf{triggering input} for $V$ if $\mathsf{Vuln}(V,x_{\mathrm{PoC}})$. An input $x$ is a \textbf{benign} (non-triggering) \textbf{input} for $V$ if $\neg\,\mathsf{Vuln}(V,x)$. The set of benign inputs is
$\mathcal{B}_V = \{\, x \mid \neg\,\mathsf{Vuln}(V,x) \,\}$.
\label{def:benign}
\end{definition}
\noindent
This definition is relative to $V$ and to the crash oracle. An input $x \in \mathcal{B}_V$ means only that $x$ produced no observed crash or sanitizer violation on $V$.

\begin{definition}
A \textbf{candidate patch} $p$ for $V$ is a proposed source-level fix; applying $p$ to $V$ yields the \textbf{patched program} $V_p$. We consider only candidates that eliminate the reported crash, i.e., $\neg\,\mathsf{Vuln}(V_p,x_{\mathrm{PoC}})$, since any other candidate is already rejected by the crash oracle.
\label{def:patch}
\end{definition}

Let $\mathsf{Obs}(P,x)$ denote the observable behavior of executing a program $P$ on an input $x$. $V$ and $V_p$ \textbf{agree} on $x$, written $V \equiv_x V_p$, if $\mathsf{Obs}(V,x) = \mathsf{Obs}(V_p,x)$, and \textbf{diverge} on $x$, written $V \not\equiv_x V_p$, otherwise. We describe the observable behavior that we compare in Section~\ref{sec:implementation}.

\begin{definition}
An input $x_w$ is a \textbf{witness input} for a candidate patch $p$ if $x_w \in \mathcal{B}_V$ and $V \not\equiv_{x_w} V_p$.
\label{def:witness}
\end{definition}

\begin{definition}
A candidate patch $p$ is \textbf{functionally divergent} if there exists a witness input for $p$.
\label{def:incorrect}
\end{definition}
\noindent
A functionally divergent patch eliminates the reported crash while changing the behavior on an input that $V$ already handles without crashing. Such a divergence indicates that the patch requires further attention from the developers. 

Establishing that no witness input exists would require checking all of $\mathcal{B}_V$. In practice, only a finite set of benign inputs can be checked.

\begin{definition}
A candidate patch $p$ is \textbf{plausible} with respect to a finite set $\mathcal{B}_{V_A} \subseteq \mathcal{B}_V$ if no input in $\mathcal{B}_{V_A} $ is a witness input for $p$:
\[
\mathrm{Valid}_{\mathcal{B}_{V_A}}(p) \;=\; \bigwedge_{a \in \mathcal{B}_{V_A}} \big(V \equiv_a V_p\big).
\]
\label{def:valid}
\end{definition}

\noindent
\textbf{Regression-freedom checking} takes $V$, $x_{\mathrm{PoC}}$, and $V_p$, and either returns a witness input or reports $p$ plausible with respect to the set $\mathcal{B}_{V_A}$ it examined. The check is one-sided: a witness input is conclusive evidence of functional divergence, whereas plausibility holds only relative to $\mathcal{B}_{V_A}$.

In the example of Section~\ref{sec:overview}, a valid in-bounds name is a benign input on which the vulnerable program creates the file and the incorrect candidate returns \texttt{NULL}; it is therefore a witness input, and the candidate is functionally divergent. The developer patch agrees with the vulnerable program on this input.
\noindent

\subsection{Intended Behavior Changes}
\label{subsec:developer_intention}
Regression-freedom checking treats the vulnerable program $V$ as the specification of behavior on benign inputs. Developers, however, sometimes change the behavior on benign inputs on purpose, for example to reject inputs that $V$ processed without crashing. Such a patch has a witness input and is therefore functionally divergent by Definition~\ref{def:incorrect}, although the divergence is intended. We observed this for developer patches in the CyberGym dataset, and present two examples.

Listing~\ref{lst:opensc-64898-vulnerable} shows a use-after-free vulnerability in \texttt{opensc/arvo\_64898}, and Listing~\ref{lst:opensc-64898-patch} shows the corresponding developer patch. \texttt{LOG\_TEST\_RET} returns from the function only when \texttt{rv} is negative. In the vulnerable code, when \texttt{rv} is 0, the \texttt{if} block frees \texttt{file}, \texttt{LOG\_TEST\_RET} does not return, and the subsequent access to \texttt{file->size} is a use after free. When \texttt{rv} is negative, the block is skipped and execution continues without crashing.
\begin{lstlisting}[
    style=deepdiffcode,
    language=C,
    caption={opensc/arvo\_64898: Vulnerable Code (Use after free when rv==0).},
    label={lst:opensc-64898-vulnerable}
]
if (!rv){ sc_file_free(file);
    LOG_TEST_RET(ctx, rv, "Get challenge error"); }
len = file->size > sizeof(buffer) ? sizeof(buffer) : file->size;
\end{lstlisting}
\begin{lstlisting}[
    style=deepdiffcode,
    language=C,
    caption={opensc/arvo\_64898: Patched Code (Changes behavior also for rv<0).},
    label={lst:opensc-64898-patch}
]
if (rv<0){ sc_file_free(file);
    LOG_TEST_RET(ctx, rv, "Get challenge error"); }
len = file->size > sizeof(buffer) ? sizeof(buffer) : file->size;
\end{lstlisting}
In the patched code, the \texttt{if} block is not entered when \texttt{rv} is 0, so \texttt{file} is not freed and the use after free is eliminated. When \texttt{rv} is negative, however, the patched code frees \texttt{file} and returns the error, whereas the vulnerable code continues. An input for which \texttt{rv} is negative is therefore a benign input on which the two programs diverge, i.e., a witness input, even though the crash occurs only when \texttt{rv} is 0. This is a developer-intended change of behavior on inputs that do not crash the vulnerable program.

Listing~\ref{lst:opensc-int-underflow-vulnerable} and Listing~\ref{lst:opensc-int-underflow-patch} show a second example, from \texttt{opensc/arvo\_18798}. The vulnerability occurs when \texttt{iACLen} equals \texttt{len}: the check \texttt{iACLen > len} does not leave the loop, and the subtraction \texttt{len -= iACLen + 1} underflows, turning the unsigned \texttt{len} into a very large value. The developer patch fixes this by changing the check to \texttt{iACLen >= len}. It also adds the condition \texttt{(size\_t)iOffset < len} to the loop, which further restricts \texttt{iOffset} and thereby changes the behavior on benign inputs as well, not only on the triggering input.
\begin{lstlisting}[
    style=deepdiffcode,
    language=C,
    caption={opensc/arvo\_18798: Vulnerable Code (Integer Underflow).},
    label={lst:opensc-int-underflow-vulnerable}
]
while (len > 1) {  /* minimum length = 2 */
    size_t iACLen = buf[iOffset] & 0x0F;
    if (iACLen > len)
        break;
    ...
    len -= iACLen + 1;
\end{lstlisting}
\begin{lstlisting}[
    style=deepdiffcode,
    language=C,
    caption={opensc/arvo\_18798: Patched Code (added restriction).},
    label={lst:opensc-int-underflow-patch}
]
while (len > 1 && (size_t)iOffset < len) {  /* minimum length = 2 */
    size_t iACLen = buf[iOffset] & 0x0F;
    if (iACLen >= len)
        break;
    ...
    len -= iACLen + 1;
\end{lstlisting}
In both examples, the developer patch, which we treat as the golden patch, is flagged as functionally divergent because it changes the behavior of the vulnerable program on benign inputs. A functionally divergent verdict should therefore be interpreted as a signal for developer review. The accompanying witness input identifies the specific input on which the behavioral change occurs, allowing developers to inspect and determine whether the change is intentional. This design favors exposing potentially unintended functional changes for developer review rather than allowing them to remain undetected during automated patch validation.

\section{Neuro-Dynamic Approaches for Regression-Freedom}
\label{sec:patchrx}
Not all candidate patches (Definition~\ref{def:patch}) are plausible patches (Definition~\ref{def:valid}). Identifying a functionally divergent patch requires determining whether the patch has changed the functional behavior of the original program on non-triggering inputs. We propose Neuro-Dynamic approaches for identifying functionally divergent patches. These approaches include dynamic analysis, neural analysis, and a hybrid combination of the two. In the following sections, we discuss how dynamic and neural techniques, individually and in combination, can support patch validation.

\subsection{Dynamic analysis based DeepDiff}
\label{subsec:dynamic_deepdiff}
In dynamic analysis-based regression-freedom checking, we execute both the vulnerable program $V$ (Definition~\ref{def:vulnprogram}) and the candidate patched program $V_p$ (Definition~\ref{def:patch}) on a set of benign inputs $\mathcal{B}_{V_A}$ that exercise the updated or modified code in the patched program. We then identify witness inputs (Definition~\ref{def:witness}) that expose functional divergence and thereby identify functionally divergent patches (Definition~\ref{def:incorrect}). Algorithm~\ref{alg:dynamic} presents the dynamic analysis approach. For each input in $\mathcal{B}_{V_A}$, the approach executes both programs, records their instrumented write traces, and compares the normalized writes with respect to the original vulnerable program. Further details on the instrumentation and normalization are provided in Section~\ref{sec:implementation}.

\begin{algorithm}[h]
\caption{Dynamic Regression-Freedom Checking}
\label{alg:dynamic}
\footnotesize
\begin{algorithmic}[1]
\Require Vulnerable program $V$, candidate patched program $V_p$ and benign inputs  $\mathcal{B}_{V_A}$
\Ensure A witness input, $x_w$
    \ForAll{$a \in \mathcal{B}_{V_A}$}
         \State $T_v[a] \gets \Call{Normalize}{\mathsf{Obs}(V,a)}$
         \State $T_p[a] \gets \Call{Normalize}{\mathsf{Obs}(V_p,a)}$
         \If{$T_v[a]
                \neq T_p[a]$}
                \State \Return $a$ 
        \EndIf
    \EndFor
\end{algorithmic}
\end{algorithm}
Since functional divergence is evaluated over a finite set of inputs $\mathcal{B}_{V_A} \subseteq \mathcal{B}V$, it is important that these inputs exercise the patched function or modified code. The benign input set $\mathcal{B}_{V_A}$ can be generated using different techniques, which can then be used by Algorithm~\ref{alg:dynamic} for regression-freedom checking.

\subsubsection{Benign input generation by mutating triggering input, $x_{\mathrm{PoC}}$.}
A naive approach to input generation is to mutate the triggering input $x_{\mathrm{PoC}}$, which reaches the vulnerable code and triggers the vulnerability. Another approach is to search for inputs adjacent to the triggering input $x_{\mathrm{PoC}}$, such as inputs at the boundary of the triggering condition that do not trigger the vulnerability. The intuition behind considering such boundary inputs~\cite{sarker2025hybrid} is that functional divergence may result from off-by-one errors or other boundary-condition changes introduced by the patch. Both approaches rely on the triggering input to generate a set of benign inputs that may exercise the updated code.

\begin{table}[t]
\footnotesize
\centering
\caption{Benign-input, $a \in \mathcal{B}_{V_A}$ generation strategies .}
\label{tab:benign-gen}
\begin{tabular}{p{2.1cm}p{8.5cm}}
\toprule
\textbf{Category} & \textbf{Strategy} \\
\midrule
Non-Neural & Mutation of $x_{\mathrm{PoC}}$ \\
Neural & Prompting with vulnerability and program context \\
Neural & Prompting with a Neuro-DeepDiff candidate witness (Section~\ref{subsubsec:neural_deepdiff}) \\
\bottomrule
\end{tabular}
\end{table}

\subsection{Neuro-Dynamic based DeepDiff}
A neural component can be integrated into regression-freedom checking in several ways. In dynamic analysis, the original and patched programs are executed on benign inputs, and their execution traces are compared. Neural analysis can complement this process at different stages:

\begin{enumerate}
\item In the purely neural approach to \DeepDiff{}, the neural component orchestrates the regression-freedom analysis and generates candidate witnesses (Definition~\ref{def:candidate-witness}) of functional divergence along with corresponding reasoning. This approach can be useful when executing the complete dynamic analysis pipeline is expensive (Section~\ref{subsubsec:neural_deepdiff}).
\item Neural analysis can assist in generating benign inputs for dynamic analysis, helping the dynamic approach exercise the modified code during patch validation (Section~\ref{subsubsec:neuro_benign_input}).
\item As discussed in Section~\ref{subsec:developer_intention}, a detected behavioral divergence may correspond to an intentional behavior change introduced by the developer. Neural analysis can assist in determining whether an observed divergence is likely to be intentional or unintended (Section~\ref{subsubsec:witness_validation}).
\end{enumerate}

\subsubsection{Neural analysis based DeepDiff (Neuro-DeepDiff)}
\label{subsubsec:neural_deepdiff}

Neuro-DeepDiff is inspired by symbolic differencing methods such as
SYMDIFF~\cite{symdiff} and UC-KLEE~\cite{ucklee}.
It checks regression freedom by comparing $V$ and $V_p$ on benign inputs
through staged, source-grounded agentic analysis, without executing the programs.
Read-only source navigation enables the agents to inspect the patch and surrounding code,
with prompts requiring evidence rather than a free-form judgment of patch correctness.
The analysis first establishes the relevant code context and input preconditions, then
enumerates non-crashing paths through changed code, and finally compares behavior along
these paths. We describe a structured path as an ordered sequence of tuples
$\langle \texttt{file}, \texttt{line}, \texttt{function}, \texttt{branch},
\texttt{assume} \rangle$, recording program locations, branch decisions, and assumptions.
The conjunction of the \texttt{assume} predicates forms the path constraint that a
candidate benign input must satisfy.
Anecdotally, we observed that without requiring this structured input-and-path
representation, the agent hallucinated purported regression-producing inputs and
corresponding pairs of program paths much more often.

The comparison phase examines benign inputs along the original crash path, without
triggering the vulnerability, as well as inputs along other non-crashing paths.
Neuro-DeepDiff aims to produce regression evidences with:
\begin{itemize}
    \item \textbf{A candidate witness (Definition~\ref{def:candidate-witness}):} an input distinct from
    $x_{\mathrm{PoC}}$, intended to exhibit the regression.
    \item \textbf{A pair of program paths:} the corresponding pre- and post-patch
    paths, with branch decisions and assumptions describing the proposed executions.
    \item \textbf{Divergence evidence:} the location and externally observable
    behavior change, with an explanation of why the original behavior was correct
    and the change is a regression rather than a repair of an existing defect.
\end{itemize}
These outputs are neuro-generated hypotheses: the proposed inputs may not realize
the reported paths, and the claimed divergence may not be reproducible in execution.
We empirically evaluate the quality of these neural artifacts against labeled
correct and incorrect patches in Section~\ref{sec:evaluation}.
Deterministic rules derive verdicts from these findings.
We repeat the analysis to gain confidence through the number of runs producing the candidate
witnesses.

We formalize these outputs and their effect on the verdict as follows.
\begin{definition}
A \textbf{candidate witness} for a candidate patched program $V_p$ is a triple $(\varphi, \pi, \delta)$, where $\varphi$ is a constraint on inputs, $\pi$ is a pair of paths through $V$ and $V_p$, and $\delta$ describes a difference in observable behavior. It claims that every input satisfying $\varphi$ is benign, follows $\pi$ in the respective programs, and exhibits $\delta$.  
A candidate witness is \textbf{confirmed} if an input derived from it is a witness input for $p$ (Definition~\ref{def:witness}), and \textbf{refuted} if an input derived from it is executed and is not a witness input for $p$.
\label{def:candidate-witness}
\end{definition}
\noindent
Here, $\varphi$ corresponds to the path constraint of the candidate input, $\pi$ to the pair of pre- and post-patch paths, and $\delta$ to the divergence evidence.

Candidate witnesses extend the notion of a plausible patch (Definition~\ref{def:valid}). Let $\mathcal{B}_{V_A} \subseteq \mathcal{B}_V$ be the finite set of benign inputs examined by execution, and let $\mathcal{B}_{V_C}$ be the set of reported candidate witnesses that are not refuted.
\begin{definition}
A candidate patch $p$ is \textbf{plausible} with respect to $(\mathcal{B}_{V_A}, \mathcal{B}_{V_C})$ if no input in $\mathcal{B}_{V_A}$ is a witness input for $p$ and $\mathcal{B}_{V_C}$ is empty:
\[
\mathrm{Valid}_{{\mathcal{B}_{V_A}},\mathcal{B}_{V_C}}(p) \;=\; \mathrm{Valid}_{\mathcal{B}_{V_A}}(p) \;\land\; \big(\mathcal{B}_{V_C} = \emptyset\big).
\]
\label{def:valid-neural}
\end{definition}
\noindent
We call $V_p$ suspected divergent if no input in $\mathcal{B}_{V_A}$ is a witness input for $V_p$ but $\mathcal{B}_{V_C}$ is not empty. Dynamic analysis reports no candidate witnesses ($\mathcal{B}_{V_C} = \emptyset$), so this definition coincides with Definition~\ref{def:valid}. Neuro-DeepDiff executes no inputs ($\mathcal{B}_{V_A} = \emptyset$) and thus we can say as suspected functional divergence. 

\subsubsection{Neural-analysis based benign input generation}
\label{subsubsec:neuro_benign_input} 
Neural analysis can also be used to generate benign inputs for regression-freedom checking. Given the vulnerability, patched code, and surrounding program context, a neural component can generate inputs that exercise the modified code without triggering the vulnerability. We consider two approaches for benign input generation using a neural component.

\paragraph{Vulnerability and Program context based Input Generation.} Benign inputs can be generated using a large language model or agent given contextual information about the vulnerability and the program. The input to the neural component can include the triggering input harness, patched code, triggering input, and crash report. Based on this context, the neural component generates benign inputs that can reach the patched code without triggering the vulnerability.

\paragraph{\textit{Neuro-DeepDiff} (Section~\ref{subsubsec:neural_deepdiff}) based Input Generation.} Given a candidate witness (Definition~\ref{def:candidate-witness}) identified by Neuro-DeepDiff, a large language model or agent is asked to generate a witness input $x_w$. This approach also uses neural analysis for input generation, but differs from the previous approach in the source of its context. Rather than directly providing the vulnerability and program context, it uses the witness description generated by \textit{Neuro-DeepDiff} to guide the generation of the witness input.
\subsubsection{Dynamic Analysis with Neuro-DeepDiff-Generated Witnesses}
\label{subsubsec:dynamic_neuro_deepdiff}

In the Neuro-DeepDiff approach, a neural component generates candidate witnesses of regression given the vulnerable and patched programs. A separate neural agent then converts these candidate witnesses into benign inputs that can be used to execute the programs. Since the candidate witnesses are generated for a specific patch, the resulting benign inputs can be used in two different ways for dynamic regression-freedom checking.

First, the benign inputs derived from the witnesses of each patch can be used only to evaluate the corresponding patch. We refer to this approach as \emph{neuro-witness-corresponding} dynamic analysis. In this approach, each patch is evaluated only with the benign inputs generated from its own candidate witnesses. This limits the dynamic analysis to the corresponding patch and its generated inputs, making it less expensive.

Second, the benign inputs generated from the witnesses of all patches can be used to evaluate all candidate patches. We refer to this approach as \emph{neuro-witness} dynamic analysis. In this approach, the benign inputs generated for one patch can also be used to evaluate other patches. Although this provides broader input coverage, it can be computationally expensive because each generated benign input may need to be executed against multiple candidate patches. In particular, when multiple witnesses are generated for each patch, the number of dynamic executions can grow substantially.

For both approaches, the generated benign inputs are executed on the vulnerable and patched programs to determine whether the candidate witnesses correspond to actual behavioral divergences. Thus, Neuro-DeepDiff-generated witnesses can support dynamic regression-freedom checking through either patch-specific execution or cross-patch reuse of witness-derived benign inputs.

\subsubsection{Neural Analysis for Understanding Intended vs. Unintended Regression}
\label{subsubsec:witness_validation}
Regression-freedom checking cannot distinguish between intended and unintended functional divergence (Section~\ref{subsec:developer_intention}). A behavioral change detected on a benign input may represent an intentional change introduced by the developer. Thus, neural analysis can provide additional context for understanding whether an observed behavioral change represents an unintended functional regression or an intentional modification by the developer. In particular, neural analysis can examine the code context, patch, and witness input to assess the nature of the observed divergence. We consider several ways in which neural analysis can provide evidence for validating an observed divergence:
\begin{itemize}
\item Multiple candidate witness inputs (Definition~\ref{def:candidate-witness} in Neuro-DeepDiff) can provide higher confidence of intentional functional divergence.
\item A candidate witness accompanied by divergence evidence of an observable behavioral change ($\delta$)  can provide a \textit{\textbf{strong}} explanation of the functional impact introduced by the patch.

\item A separate neural component can act as a validator to assess whether the observed divergence identified by dynamic analysis is more likely to result from an unintended functional change or an intentional behavior modification.
\end{itemize}

\section{Implementation}
\label{sec:implementation}
In this section, we discuss the key implementation details of our \DeepDiff{} approach.

\paragraph{Write traces as observable behavior.}
For observable behavior comparison, we capture writes  during execution. We instrument write operations and record the corresponding left-hand-side variables after each write. The instrumentation is heuristic-based and does not perform data-flow or semantic analysis.

\paragraph{Instrumented function.}
Instrumenting writes throughout the entire codebase can be expensive. Since a patch typically modifies a small portion of the codebase, we identify the function affected by the vulnerability or patch using information from the crash report and/or the \texttt{git diff}, and instrument only the affected function.

\paragraph{Trace normalization.}
Before comparing write traces in Algorithm~\ref{alg:dynamic}, we normalize the recorded writes. Memory addresses can vary across builds and executions, and directly comparing them can therefore introduce spurious divergences. We use heuristics to identify memory-related writes, including writes involving pointer dereferences and writes whose right-hand sides contain operations such as \texttt{memcpy}, \texttt{malloc}, or \texttt{alloc}. We also treat different orderings of the same writes as equivalent, since reordering writes does not necessarily indicate behavioral divergence.

\paragraph{Filtering benign inputs.}
We filter generated inputs before considering them benign inputs according to Definition~\ref{def:benign}. Each generated input is executed on the original vulnerable program, and inputs that trigger a crash according to the sanitizer oracle are discarded.

\paragraph{Reachability for regression-freedom.}
While recording writes, we also track whether a benign input reaches the vulnerable or patched function and whether it reaches the code modified by the patch. This reachability information is used to determine whether an input exercises the code relevant to regression-freedom checking.

\paragraph{Mutation-based benign input generation.}
Our mutation-based input generation first applies deterministic, structure-preserving mutations. Specifically, it systematically replaces each adjacent byte pair in the triggering input with small values encoded in both big- and little-endian order. It then falls back to randomized byte-level transformations. We specify both a target number of benign inputs and an upper bound on the number of mutations. If the target cannot be reached within this bound, the approach may produce fewer benign inputs or none. For boundary-based mutation, we generate inputs adjacent to the triggering input to explore boundary conditions around the vulnerability-triggering input.

\paragraph{Candidate witness inputs.}
Candidate witnesses (Definition~\ref{def:candidate-witness}) generated by Neuro-DeepDiff are represented in a structured JSON format.

\paragraph{LLM models used to generate benign inputs.} For our neural analysis, we use Claude Opus 4.6~\cite{claude_opus} as the agentic model for generating benign inputs and for our Neuro-DeepDiff approach. We selected this model for its ability to operate on and reason over large codebases.

\section{Experimental Evaluation}
\label{sec:evaluation}
We derive our evaluation corpus from CyberGym-E2E~\cite{DBLP:journals/corr/abs-2606-04460}, a benchmark constructed from historical vulnerabilities reported by OSS~\cite{oss_fuzz}. CyberGym-E2E associates each vulnerability with a reproducible vulnerable program, the corresponding developer patch, a triggering input, and a containerized build environment. Its construction pipeline retains an instance only when the project builds successfully and the triggering input triggers the target vulnerability in the vulnerable program but not in the patched version. We select a subset of these instances in which the vulnerability and the corresponding patched code are located within the same function and the developer patch contains one to three hunks of code changes to construct our evaluation corpus for regression-freedom checking.

Our corpus contains 630 patches associated with 118 vulnerabilities from 40 open-source projects. CyberGym-E2E focuses on memory-safety vulnerabilities, and the vulnerabilities in our extracted corpus include out-of-bounds accesses, uninitialized memory, heap/stack overflow, and integer overflow and underflow. Each vulnerability contributes one developer-written patch, and we generate additional patches by mutating the corresponding developer-written patch.

\subsection{Incorrect patch generation by mutating developer patch}
Evaluating a patch validator requires incorrect patches that are semantically non-equivalent to the developer patches. CyberGym-E2E provides only the developer patch for each vulnerability. Therefore, we synthetically generate incorrect patches following the approach of OmniCode~\cite{sonwane2026omnicode}, which introduces commonly occurring patching errors. However, unlike OmniCode, which uses agents to generate incorrect patches through mutation-based transformations, we directly apply mutation operators to the developer-written patches provided by CyberGym-E2E.

Our mutation process applies between one and three localized semantic transformations to each developer-written patch. The mutation operators include changing arithmetic or relational operators, swapping conjunctions and disjunctions, negating conditions, introducing off-by-one changes, swapping function parameters, and replacing return values. These transformations are designed to model plausible patching errors while avoiding intentionally malformed patches. The mutation manifest classifies all generated mutants as semantic rather than syntactic mutations.

Depending on the availability of applicable mutation sites, each vulnerability contributes between one and seven patches. Across the complete dataset, each vulnerability contributes an average of 5.34 patches, including 4.34 mutated patches. Of the 630 patches, 496 are semantically mutated patches, while the remaining 134 comprise the developer-written patches and patches that are semantically equivalent to the developer-written patches.

\subsection{Dataset for Regression-Freedom Analysis}

Of the 496 semantically mutated patches, 245 fail to eliminate the vulnerability because the triggering input still causes a crash in the patched program. We exclude these patches from our regression-freedom analysis because they do not satisfy the basic requirement of a candidate patch (Definition~\ref{def:patch}). This filtering leaves 385 candidate patches, comprising 251 semantically mutated patches and 134 patches that are developer-written or semantically equivalent to them. The patches contain one to three hunks of code changes, with an average of 9.01 changed lines per patch. We use these 385 candidate patches to evaluate our Neuro-Dynamic regression-freedom checking approach, \DeepDiff{}.

\subsection{Metrics Used for Evaluation}

We consider a candidate patch to be incorrect if it is semantically non-equivalent to the developer-written patch. We treat the identification of incorrect patches as the positive class. For evaluating the regression-freedom approaches, we focus on three primary metrics:

\begin{enumerate}
\item \textbf{Recall:} Recall measures the proportion of incorrect patches identified by the approach out of all incorrect patches in the evaluation corpus. In our dataset, there are 251 incorrect (semantically mutated) patches.
\item \textbf{False Alarm:} False alarm measures the proportion of patches that are incorrectly flagged as functionally divergent among the developer-written or semantically equivalent patches. It is calculated as the number of such patches flagged as functionally divergent divided by the total, which is 134.

\item \textbf{F1 Score (Average):} F1 score is the harmonic mean of the precision and recall. Since our dataset is imbalanced, the overall F1 score can be dominated by the majority class. Therefore, we report the average F1 score, which is calculated as the arithmetic mean of the F1 scores for the positive and negative classes, giving equal weight to both.
\end{enumerate}

\subsection{Research Questions}

Our experimental evaluation targets the following research questions:

\textbf{RQ1.} To what extent does regression-freedom-based patch validation improve the detection of incorrect patches?

\textbf{RQ2.} How do different strategies for generating benign inputs affect the effectiveness of the patch validation pipeline?

\textbf{RQ3.} How effective is neural analysis in distinguishing intended behavior changes from unintended functional regressions?

\textbf{RQ4.} How effective is \textit{Neuro-DeepDiff} for patch validation?

\subsection{Experimental Results}
\subsubsection{RQ1: Effectiveness of Regression-Freedom Checking}

\begin{figure}[t]
    \centering
    \includegraphics[width=0.5\linewidth]{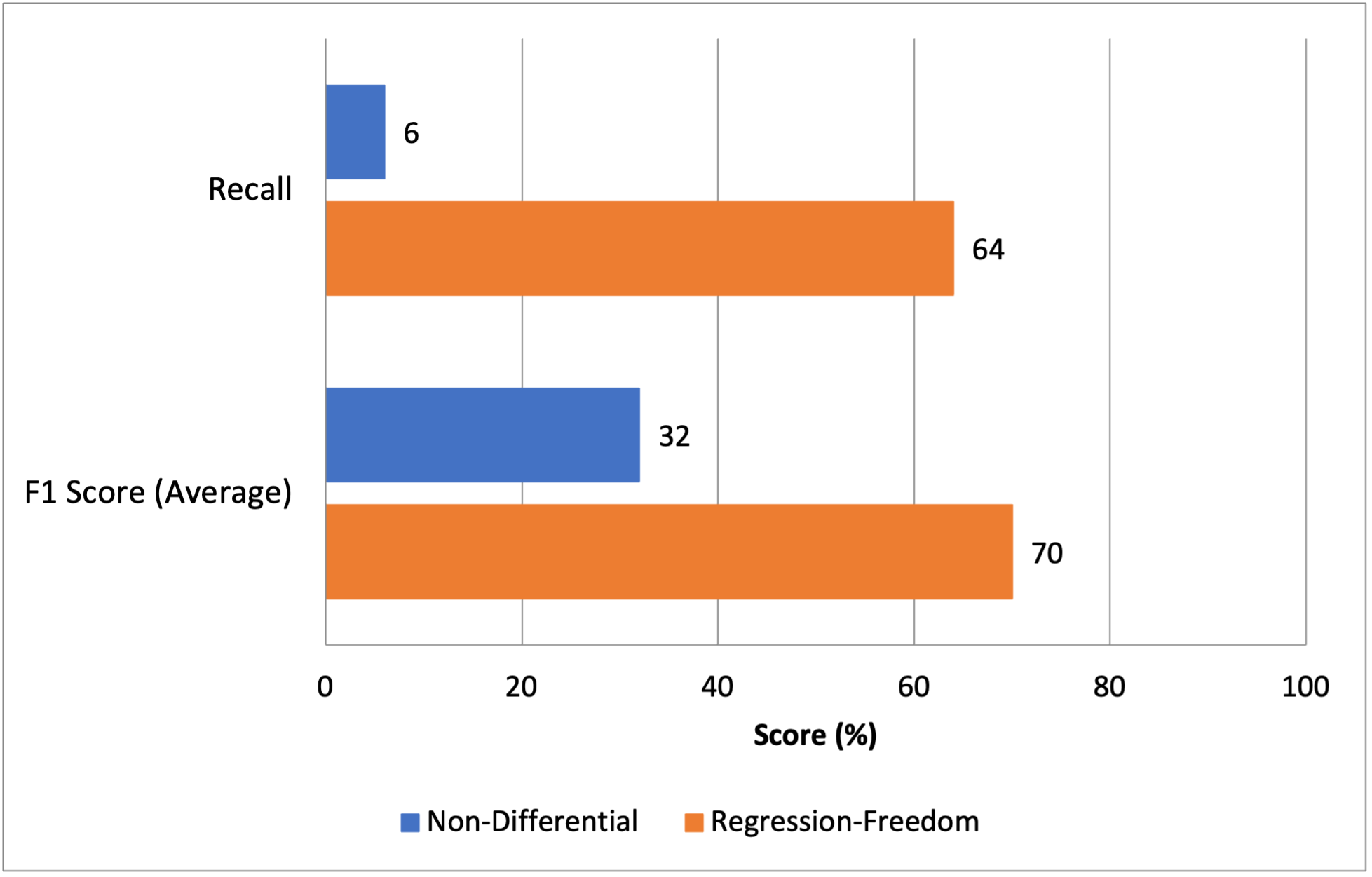}
    \caption{Effectiveness of Regression-Freedom checking}
    \label{fig:diff_no_diff_analysis}
\end{figure}

To evaluate the effectiveness of regression-freedom checking for security patch validation, we investigate whether simply executing the vulnerable and patched programs on benign inputs is sufficient to identify incorrect patches, or whether differential analysis is necessary to expose functional divergence. We therefore compare a non-differential approach, which executes the programs on benign inputs without comparing their behavior, with our regression-freedom checking approach, which differentially compares their behavior. Fig.~\ref{fig:diff_no_diff_analysis} presents the recall and average F1 score of the non-differential and regression-freedom approaches.

The non-differential approach identifies only 6\% of the incorrect patches, whereas regression-freedom checking identifies 64\%, representing a 58 percentage-point improvement in recall. In the non-differential approach, a patch is flagged as incorrect if executing it on a benign input triggers a new vulnerability or results in a build failure. In contrast, regression-freedom checking can identify functional divergence by comparing the behavior of the vulnerable and patched programs. This improvement, however, comes with an increase in false alarms. Regression-freedom checking flags 18\% of the developer-written or semantically equivalent patches as functionally divergent. This occurs because regression-freedom checking identifies behavioral differences on benign inputs without distinguishing whether those changes are intentional modifications introduced by developers or unintended functional regressions.

Despite this trade-off, regression-freedom checking improves the average F1 score by 38 percentage points compared with the non-differential approach. These results indicate that comparing the behavior of the vulnerable and patched programs provides substantially more information for identifying functionally divergent patches than simply executing the programs on benign inputs.

Overall, there is a trade-off between the detection of incorrect patches and having false alarms. Even though we are improving recall by 58\%, the 18\% false alarm has been introduced in our regression-freedom analysis approach. Our regression-freedom analysis looks for functional divergence based on benign inputs which do not trigger the vulnerability in the vulnerable code. Thus, it might flag developers' intended change also as functional divergence. However, the F1 score shows that overall, regression-freedom checking improve the performance by 38\%.

\subsubsection{RQ2: Effectiveness of Benign Input Generation Approaches}
\begin{figure}[t]
    \centering
    \begin{subfigure}{0.48\linewidth}
        \centering
        \includegraphics[width=\linewidth]{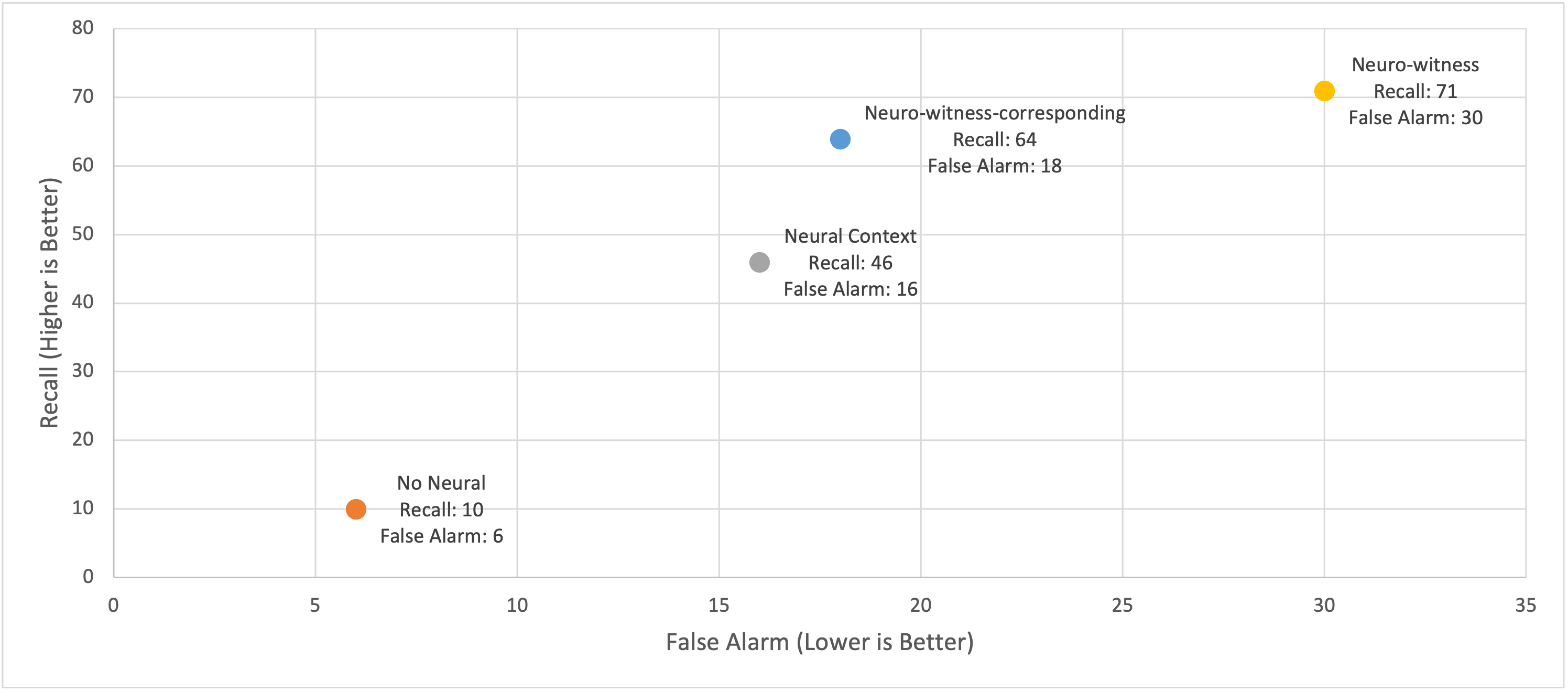}
        \caption{Recall vs False Alarm}
        \label{fig:benign_input_effectiveness_recall_fa}
    \end{subfigure}
    \hfill
    \begin{subfigure}{0.48\linewidth}
        \centering
        \includegraphics[width=\linewidth]{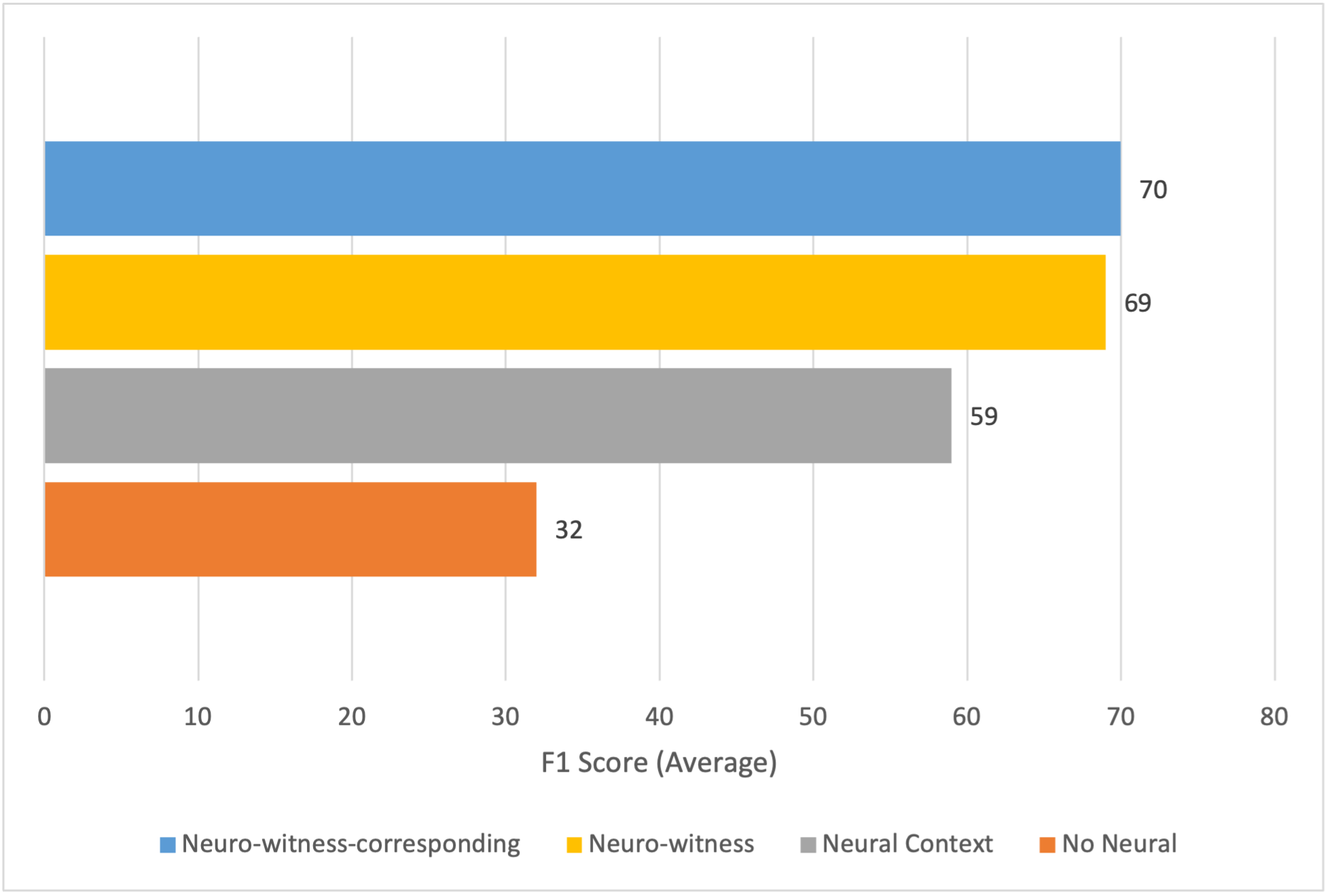}
        \caption{Average F1 score}
        \label{fig:benign_input_effectiveness_f1}
    \end{subfigure}
    \caption{Effectiveness of different benign input generation approaches.}
    \label{fig:benign_input_effectiveness}
\end{figure}

Fig.~\ref{fig:benign_input_effectiveness}(a) presents the recall and false alarm rate for different benign input generation approaches. The inputs generated without neural analysis have the lowest recall, at 10\%, while producing a false alarm rate of 6\%. The other three input generation approaches use neural approaches. The neural context approach corresponds to vulnerability and program context based input generation, both neuro-witness and neuro-witness-corresponding are Neuro-DeepDiff based input generation (Discussed in Section~\ref{subsubsec:neuro_benign_input}).

As shown in Fig.~\ref{fig:benign_input_effectiveness}(a), the neural-based input generation approaches achieve substantially higher recall than the approach without neural assistance. The neuro-witness approach achieves the highest recall of 71\%, but also has the highest false alarm rate of 30\%. The neuro-witness-corresponding approach achieves a recall of 64\% with a false alarm rate of 18\%, while the neural context achieves a recall of 46\% with a false alarm rate of 16\%. These results indicate that witnesses derived using Neuro-DeepDiff can expose more functionally divergent patches than either non-neural or context based input generation approaches. However, the results also demonstrate a trade-off between identifying incorrect patches and introducing false alarms.

Fig.~\ref{fig:benign_input_effectiveness}(b) presents the average F1 scores of the different benign input generation approaches. The neuro-witness-corresponding approach achieves the highest F1 score, indicating the best balance between identifying incorrect patches and limiting false alarms among the evaluated approaches. The neuro-witness approach achieves the second-highest F1 score. The neural context approach achieves an F1 score of 59\%, which is 11 percentage points lower than the neuro-witness-corresponding approach and 10 percentage points lower than the neuro-witness one. The approach without neural assistance achieves the lowest F1 score of 32\%. Overall, these results indicate that Neuro-DeepDiff witness-derived benign inputs, particularly when used for the corresponding patch, provide an effective strategy for regression-freedom checking in security patch validation.

\subsubsection{RQ3: Effectiveness of Neural Analysis for Understanding Intended Behavior Changes}
\begin{figure}[t]
    \centering
    \begin{subfigure}{0.48\linewidth}
        \centering
        \includegraphics[width=\linewidth]{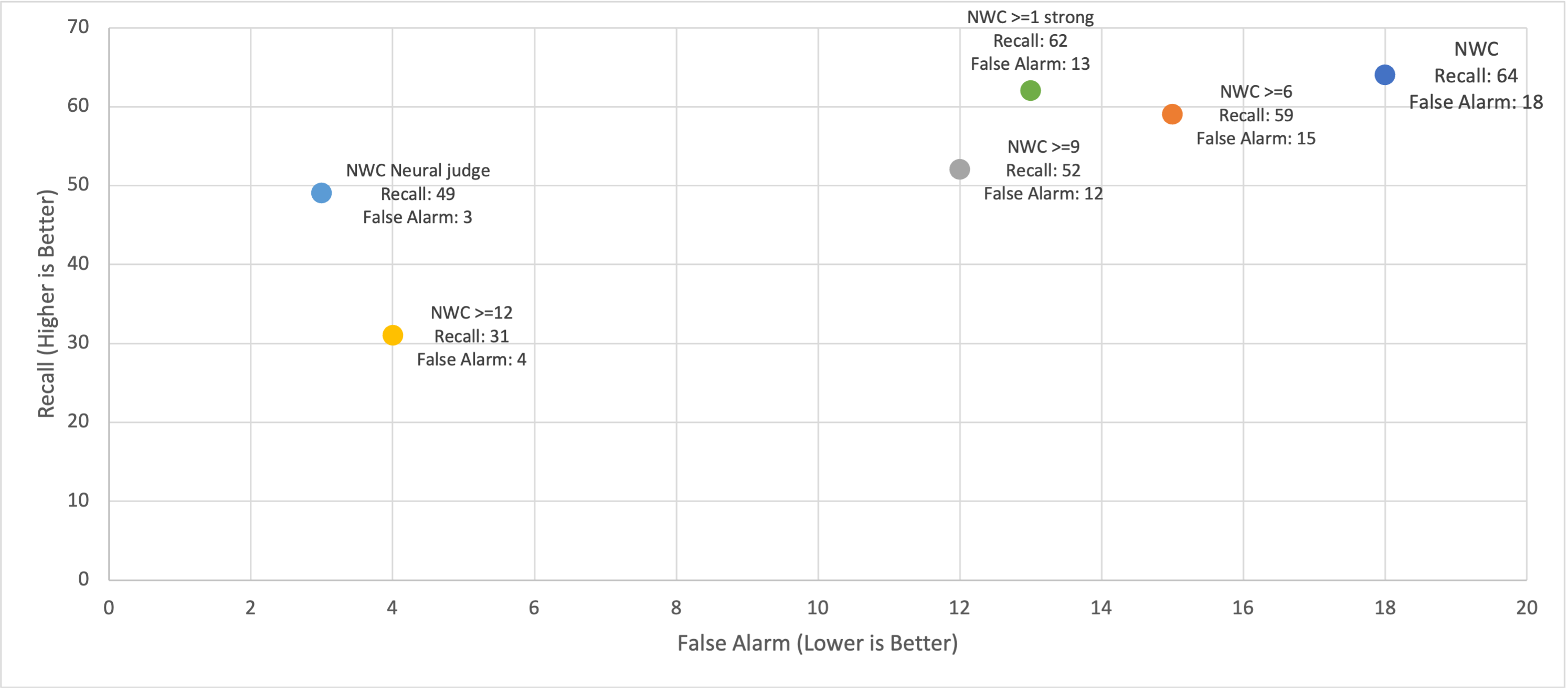}
        \caption{Recall vs False Alarm}
    \end{subfigure}
    \hfill
    \begin{subfigure}{0.48\linewidth}
        \centering
        \includegraphics[width=\linewidth]{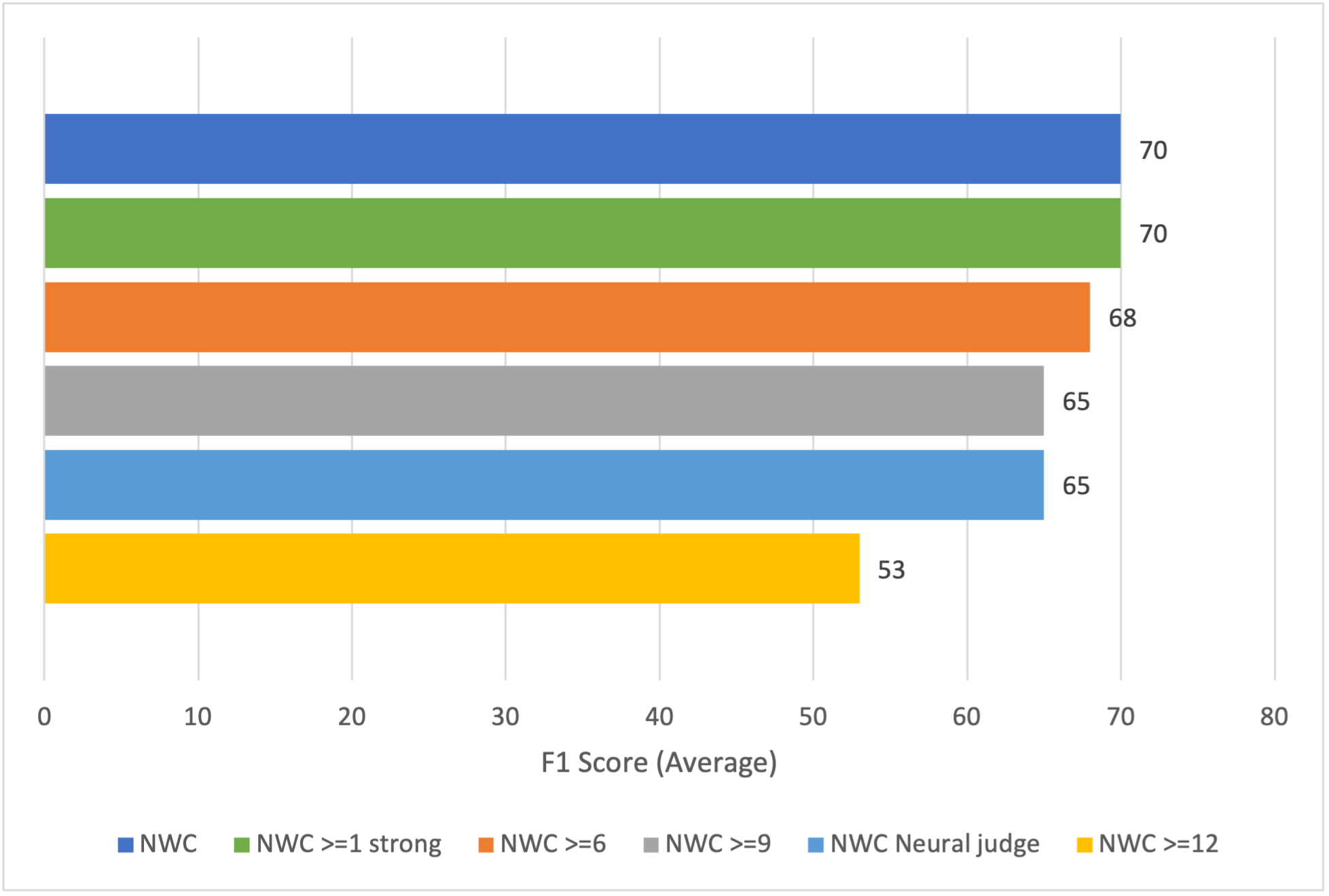}
        \caption{Average F1 score}
    \end{subfigure}
    \caption{Effectiveness of the Neural Analysis for intended Regression Reasoning.}
    \label{fig:agent_validator_effectiveness}
\end{figure}
Dynamic regression-freedom analysis can detect behavioral changes that may have been intentionally introduced by developers. Neural analysis can provide additional evidence for determining whether an observed divergence is more likely to represent an intentional behavior change or an unintended functional regression. In this research question, we investigate the effectiveness of different neural validation strategies for analyzing detected regressions.

As RQ2 shows that the neuro-witness-corresponding (NWC) approach achieves the highest F1 score among the benign input generation approaches, we evaluate the different neural strategies using NWC. In NWC, benign inputs are generated from the candidate witnesses produced by Neuro-DeepDiff and then dynamically executed on the corresponding vulnerable and patched programs. We investigate whether having multiple candidate witnesses or a strong divergence evidence from a candidate witness accompanied by an observable behavioral change ($\delta$) (Section~\ref{subsubsec:witness_validation}) can reduce false alarms while retaining the ability to identify incorrect patches.

Fig.~\ref{fig:agent_validator_effectiveness}(a) presents the recall and false alarm rate of the different validation strategies. We consider NWC $\geq N$, where $N$ represents the number of candidate witnesses generated by neuro-deepdiff, which requires at least $N$ candidate witnesses to flag a patch incorrect. We also consider NWC $\geq 1$ strong, which requires at least one candidate witness with divergence evidence ($\delta$). Finally, NWC with neural judge uses an independent neural validator to evaluate the observed regression based on the dynamic analysis results.

The NWC approach without additional validation achieves the highest recall of 64\%, with a false alarm rate of 18\%. NWC $\geq 1$ strong achieves the second-highest recall of 62\%, while reducing the false alarm rate to 13\%. For NWC $\geq N$, both recall and false alarm rate decrease as the required number of witnesses increases. This indicates that requiring multiple witnesses provides additional evidence for a behavioral divergence and can reduce false alarms, but this filtering comes at the cost of lower recall.

Fig.~\ref{fig:agent_validator_effectiveness}(b) presents the F1 scores of the different validation strategies. NWC and NWC $\geq 1$ strong achieve the highest F1 scores. In particular, requiring at least one strong witness reduces recall only slightly, from 64\% to 62\%, while reducing the false alarm rate from 18\% to 13\%. These results suggest that using strong witness evidence with the neuro-witness-corresponding approach can provide a better balance between identifying incorrect patches and reducing false alarms. The strong-witness criterion also provides additional evidence that can help developers assess whether an observed regression may represent an intentional behavior change rather than an unintended functional regression.

\subsubsection{RQ4: Effectiveness of the Neuro-DeepDiff Approach}
\begin{figure}[t]
    \centering
    \begin{subfigure}{0.48\linewidth}
        \centering
        \includegraphics[width=\linewidth]{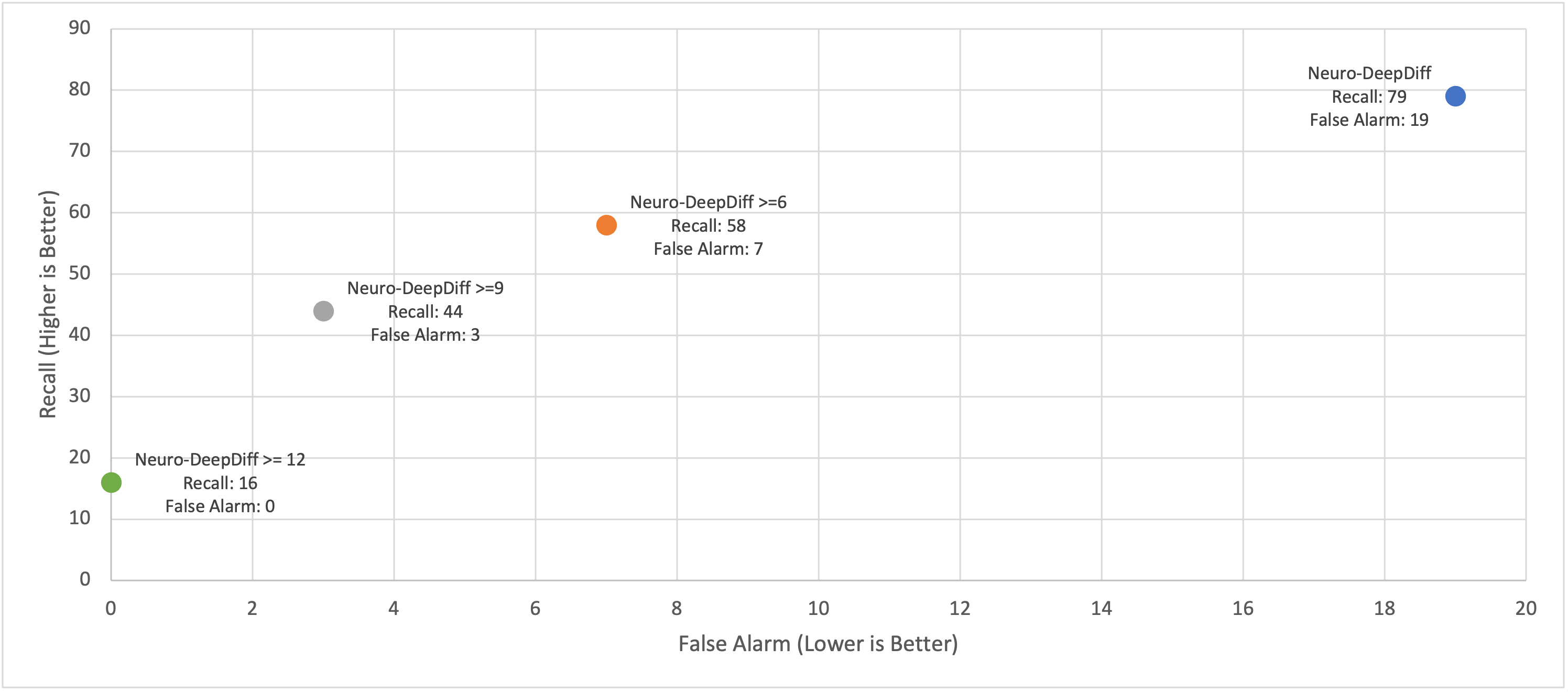}
        \caption{Recall vs False Alarm}

    \end{subfigure}
    \hfill
    \begin{subfigure}{0.48\linewidth}
        \centering
        \includegraphics[width=\linewidth]{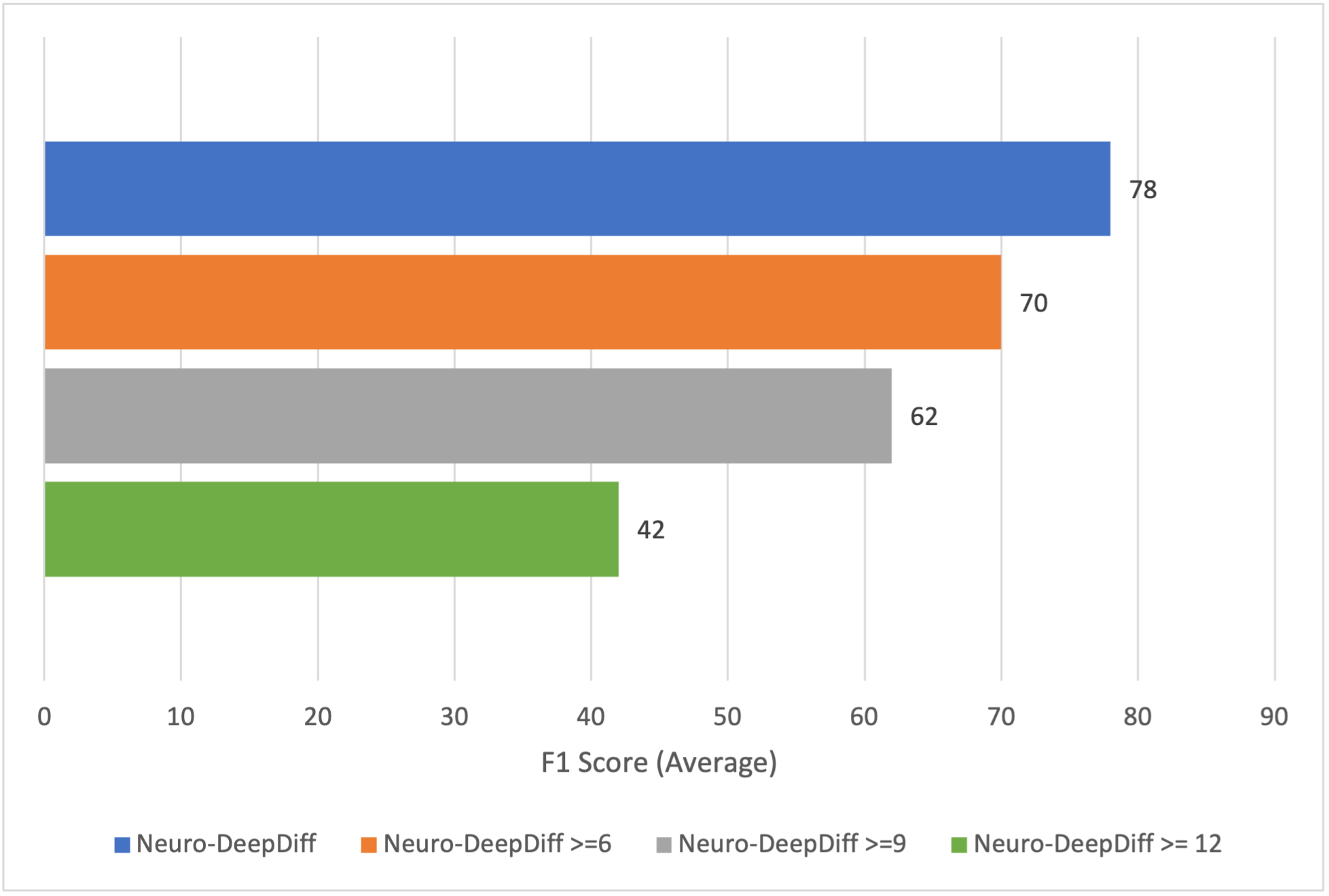}
        \caption{Average F1 Score}
       
    \end{subfigure}
    \caption{Effectiveness of Neuro-\DeepDiff{} approach.}
    \label{fig:neuro_deepdiff_effectiveness}
\end{figure}

In this research question, we investigate the effectiveness of the Neuro-DeepDiff approach. This evaluation is important because it may not always be feasible to dynamically execute candidate patches with benign inputs due to the effort required to set up and maintain the execution environment. We therefore evaluate Neuro-DeepDiff based on the number of candidate witnesses generated with divergence evidence.

Figure~\ref{fig:neuro_deepdiff_effectiveness} presents the effectiveness of Neuro-DeepDiff based on the number of candidate witnesses generated with divergence evidence. For Neuro-DeepDiff, a patch is considered incorrect if at least one candidate witness is generated with divergence evidence. Similarly, for Neuro-DeepDiff $\geq N$, a patch is considered incorrect if at least $N$ candidate witnesses are generated with divergence evidence. We observe a similar trend to that in RQ3, where increasing the required number of candidate witnesses reduces both recall and false alarms. Neuro-DeepDiff with at least one candidate witness achieves the highest recall of 79\%, with a false alarm rate of 19\%. In contrast, Neuro-DeepDiff $\geq 12$ achieves the lowest recall of 16\%, with no false alarms.

The F1 score follows a similar trend. Neuro-DeepDiff with at least one candidate witness achieves the highest F1 score of 78\%, while Neuro-DeepDiff $\geq 6$ achieves the second-highest F1 score of 70\%. The F1 score further decreases to 62\% and 42\% for Neuro-DeepDiff $\geq 9$ and Neuro-DeepDiff $\geq 12$, respectively. These results indicate that requiring more candidate witnesses with divergence evidence provides more conservative patch validation and reduces false alarms, but also reduces the ability to identify incorrect patches.

We further investigate whether the candidate witnesses generated by Neuro-DeepDiff can be converted into executable benign inputs for dynamic regression-freedom checking. Neuro-DeepDiff generates a total of 3039 candidate witnesses across the evaluation dataset, of which 3037 are converted into inputs by a neural component. We then filter these inputs to retain only those that do not trigger any vulnerability in the original vulnerable program, according to Definition~\ref{def:benign}. Of the 3037 converted inputs, 2898 satisfy this criterion, corresponding to 95\% of the converted inputs. We then execute these benign inputs on the corresponding vulnerable and patched programs using dynamic analysis. Dynamic analysis identifies behavioral divergence for 1876 of the 2898 benign inputs. Thus, in approximately 65\% of cases, the dynamically identified witnesses $x_w$ align with the candidate witnesses $\mathcal{B}_{V_C}$ generated by Neuro-DeepDiff.

Overall, these results demonstrate that Neuro-DeepDiff can effectively generate candidate witnesses with divergence evidence for security patch validation. Moreover, the high proportion of generated inputs that satisfy the benign-input criterion and the 65\% agreement between candidate witnesses and dynamically identified witnesses indicate that Neuro-DeepDiff-generated witnesses can provide useful evidence of functional divergence without requiring dynamic execution during the initial witness-generation process.

\subsection{Discussion}

Across all the research questions, we observe a trade-off between the recall of identifying incorrect (semantically mutated) patches and the false-alarm rate of flagging developer-written or semantically equivalent patches as incorrect. As the number of benign inputs increases, more execution paths are explored, increasing the likelihood of reaching edge cases. Consequently, while generating more benign inputs can improve the detection of incorrect patches, it can also lead to an increase in false alarms. We also found that, in some cases, developers intentionally change the program behavior for inputs that do not trigger a vulnerability under evaluated sanitizer oracle. Although dynamic analysis based witness inputs may flag a correct developer patch as divergent, they provide witness inputs along with the corresponding divergent execution traces. We believe that this information can help developers understand the behavioral differences and make an informed decision about whether the observed divergence is intended. Furthermore, we believe that the consequences of incorrectly classifying an incorrect patch as a good patch can be more severe, as a functionally incorrect patch may be deployed and potentially leave the underlying vulnerability exploitable. In contrast, when a developer-written patch is identified as incorrect, the provided witness inputs and divergent traces can be used by developers to further investigate the reported behavior and determine whether the divergence is intentional.

\subsection{Threats to Validity}

\paragraph{Internal validity.} Our incorrect patches are generated by mutating developer-written patches (Section~\ref{sec:evaluation}) rather than being collected from independently authored incorrect patches. Although these mutations model plausible patching errors, they may not fully represent the distribution of errors that occur in real-world automated patch generation. Nevertheless, mutation is a commonly used approach for synthetically generating incorrect patches~\cite{sonwane2026omnicode}. Our regression-freedom analysis also incorporates neural components at different stages, which may introduce non-determinism. To assess the stability of our results, we ran our pipelines multiple times and observed similar results across runs. We therefore consider the reported results to be representative of the evaluated setting.

\paragraph{External validity.} Our dataset is drawn from CyberGym-E2E and is restricted to C/C++ projects instrumented by OSS-Fuzz with a sanitizer-based crash oracle. Consequently, our results may not generalize to memory-safe languages or to vulnerability classes that do not manifest as crashes, such as logic errors or authorization bypasses. Nevertheless, the concept of regression-freedom is not inherently tied to a particular programming language or crash oracle and could potentially be applied to other languages and other observable behaviors. Our neural components may also have encountered similar vulnerabilities, patches, or code patterns during pretraining. However, our approach does not rely solely on the neural components to determine patch correctness. Instead, the neural components are used to generate candidate witnesses and divergence evidence, which are subsequently used as concrete evidence for the regression-freedom analysis and require additional reasoning to establish a behavioral difference.

\subsection{Limitations} The write-trace-based regression-freedom checking approach has several limitations. For example, it may have difficulty handling cases where a variable is renamed or replaced, a variable assignment is introduced within a loop, a new variable is added, or memory addresses are compared. To mitigate some of these cases, we apply heuristics, such as ignoring comparisons involving memory addresses. Addressing other cases may require more sophisticated semantic data-flow analysis and type analysis or inference. These limitations can result in false alarms. However, our manual analysis shows that only 29\% of the false alarms are attributable to limitations of the regression analysis itself. These cases could potentially be addressed through improved heuristics for write-trace analysis.

\section{Related Works}
\label{sec:relwork}
\paragraph{Automated Patch Evaluation.}
Existing research on automated patching has primarily addressed the patch overfitting problem with respect to a given test suite~\cite{fei2025patch,petke2024patch}. Other approaches assess and evaluate patches using LLM-based judges or pretrained models~\cite{shi2026towards,zhang2026compass,li2026furina,10659742}, augmenting existing test suites with additional functional tests or vulnerability-triggering inputs~\cite{DBLP:journals/corr/abs-2606-04460,yu2026patch,jang2026improving}, or leveraging information from historical developer discussions~\cite{lin2026learning}. A separate line of research focuses on identifying security patches and silent vulnerability fixes~\cite{sawadogo2022sspcatcher,tian2012identifying,zhou2021finding,wu2022enhancing,wang2019detecting,zhou2021spi}, including approaches that prioritize patches by inferring vulnerabilities from patching information~\cite{vulDisPatchCommitFei24,wang2021patchrnn,sawadogo2022sspcatcher,tian2012identifying,ChengHUANG:52,wang2020machine,zhou2021finding,wu2022enhancing,wang2019detecting}. However, these approaches do not explicitly assess whether a security patch changes the functional behavior of the original program. Our regression-freedom checking approach addresses this gap by differentially evaluating the functional behavior of security patches against the original vulnerable program.

\paragraph{LLM-Generated Code Evaluation.}
Several works have studied the evaluation of LLM-generated code~\cite{improvingPromt,nondeterminism,programmingLangComp,xu2022systematic,jiang2022discovering,ahmed2023improving,CopilotCodewhispererChatGPT,EvalPlus}. These approaches primarily evaluate functional correctness or general code quality rather than the functional behavior regression of security patches.

\paragraph{Differential Analysis and Testing.}
Numerous differential analysis and testing approaches use symbolic execution for program equivalence analysis~\cite{ardiff,dse,IMPS,pasda,PEQcheck,PEQtest,ucklee,ModDiff,clever,symdiff,rvt,shadowSym,shadowSym2,shadowSym3,symbolicExecution}. Differential fuzzing~\cite{inMemoryFuzzing,fuzzDiff,PEM} and differential monitoring~\cite{diffMonitoring} provide scalable alternatives for identifying behavioral differences, while hybrid approaches combine differential symbolic execution and fuzzing~\cite{sarker2025hybrid}. However, these approaches do not focus on differential patch validation using the original vulnerable program as the behavioral reference.

\paragraph{Regression Verification.}
Prior work has also explored static differential verification to determine whether bug fixes introduce regressions~\cite{lahiri2013differential,logozzo2014verification}. These approaches formulate regression checking in terms of program assertions and verify that the assertions remain satisfied after a program modification. Our notion of regression freedom instead identifies behavioral differences between the original and patched programs on non-triggering inputs without requiring additional program assertions to specify the expected behavior.

\section{Conclusion And Future Directions}
\label{sec:conclusion}
Passing an existing test suite and no longer crashing on a triggering input does not establish that a security patch is correct: neither check observes behavior on inputs that the test suite never exercises, allowing a functionally divergent patch to satisfy both while silently changing program behavior. We introduced regression-freedom checking, a validation criterion that treats the original vulnerable program as the behavioral reference and uses benign inputs to expose such divergences. We instantiated this criterion through dynamic and neural analysis and extended the existing CyberGym dataset with semantically mutated incorrect patches, resulting in a total of 630 patches, including 496 incorrect patches. For the same set of benign inputs, regression-freedom checking substantially improves recall over non-differential analysis, increasing it from 6\% to 64\%. Our analysis also demonstrates that functional divergence can be intentional, highlighting the challenge of distinguishing unintended regressions from intended behavior changes.

Several directions remain for future work. Developer-intent analysis~\cite{sharma2026contextcov,haing2026devintent} could provide additional evidence for distinguishing unintended regressions from intentional behavior changes. Moreover, moving from benign-input-based analysis toward benign path based analysis, potentially using symbolic execution~\cite{symbolicExecution}, could provide more systematic coverage of non-crashing program behaviors and increase confidence in regression-freedom checking. Our current approach also focuses primarily on regression testing and does not explicitly investigate the generation of multiple vulnerability-triggering inputs. Generating a broader set of triggering inputs before benign-input generation could improve coverage of the vulnerability-triggering behavior and provide a stronger foundation for subsequent regression-freedom checking.

\bibliographystyle{ACM-Reference-Format}
\bibliography{main}

\end{document}